\documentclass[aps,twocolumn,superscriptaddress,showpacs,floatfix,preprintnumbers]{revtex4-1}

\usepackage{graphicx} 
\usepackage{float}
\usepackage{dcolumn}
\usepackage{bm}
\usepackage{amsmath,xcolor,comment}
\newcommand{\data}{{\rm data}}
\newcommand{\EoS}{{\rm EoS}}

\begin{document}

\title{Astrophysical Equation-of-State Constraints on the Color-Superconducting Gap}
\author{Aleksi Kurkela}%
\affiliation{Faculty of Science and Technology, University of Stavanger, 4036 Stavanger, Norway}
\author{Krishna Rajagopal}
\affiliation{Center for Theoretical Physics, Massachusetts Institute of Technology, Cambridge, MA  02139, USA}
\affiliation{Theoretical Physics Department, CERN, 1211 Geneva 23, Switzerland}
\author{Rachel Steinhorst}
\affiliation{Center for Theoretical Physics, Massachusetts Institute of Technology, Cambridge, MA  02139, USA}

\date{April 2024}

\preprint{MIT-CTP/5671}

\begin{abstract}
We demonstrate that astrophysical constraints on the dense-matter equation of state place an upper bound on the color-superconducting gap in dense matter above the transition from nuclear matter to quark matter. Pairing effects in the color-flavor locked (CFL) quark matter phase increase the pressure at high density, and if this effect is sufficiently large then the requirements of causality and mechanical stability make it impossible to reach such a pressure in a way that is consistent with what is known at lower densities. The intermediate-density equation of state is inferred by considering extensions of chiral effective field theory (CEFT) to neutron star densities, and conditioning these using current astrophysical observations of neutron star radius, maximum mass, and tidal deformability (PSR J0348+0432, PSR J1624-2230, PSR J0740+6620, GW170817). At baryon number chemical potential $\mu = 2.6~\text{GeV}$ we find a 95\% upper limit on the CFL pairing gap $\Delta$ of $457~\text{MeV}$ using overly conservative assumptions and 
$216~\text{MeV}$ with more reasonable assumptions. This constraint may be strengthened by future astrophysical measurements as well as by future advances in high density QCD calculations.
\end{abstract}

\maketitle

\section{Introduction}

Observations of pulsars~\cite{Demorest:2010bx,Antoniadis:2013pzd,NANOGrav:2019jur,Fonseca:2016tux,Fonseca:2021wxt,Riley:2019yda, Miller:2019cac,Riley:2021pdl,Miller:2021qha} and neutron-star 
mergers~\cite{LIGOScientific:2017vwq, LIGOScientific:2018cki,LIGOScientific:2018hze,LIGOScientific:2020aai, LIGOScientific:2017zic}
place constraints on the thermodynamic properties of neutron star matter, allowing the inference of the equation of state (EoS) at neutron star densities~
\cite{Read:2008iy,Lindblom:2010bb,Lindblom:2012zi,Hebeler:2013nza, Kurkela:2014vha,Annala:2017llu,Tews:2018iwm,Landry:2018prl,Greif:2018njt, Most:2018hfd, LIGOScientific:2018cki, Annala:2019puf,Raaijmakers:2019dks,Essick:2020flb,Annala:2021gom,Altiparmak:2022bke,Jiang:2022tps,Gorda:2022jvk,Annala:2023cwx}. 
Densities inside neutron stars ($n \lesssim (5 - 8) n_s$, with $n_s\approx 0.16~{\rm fm}^{-3}$ being the density of ordinary nuclear matter) are not high enough to be described in a controlled fashion by the rigorous calculations of perturbative quantum chromodynamics (pQCD). It has nevertheless been demonstrated that knowledge of the pQCD EoS at higher densities ($n \sim (20-40) n_s$), where the calculation {\it is} under analytic control, places an additional set of non-trivial constraints (arising from the requirements of mechanical stability, causality and thermodynamic consistency)
on the EoS at neutron-star densities as well as on integrals of the EoS between neutron star densities and the high pQCD-densities~\cite{Komoltsev:2021jzg, Gorda:2022jvk,Komoltsev:2023zor}.
This interdependence between the neutron-star EoS and the high-density EoS also allows us to ask the converse question: can astrophysical observations of neutron stars be used to obtain empirical access to high-density QCD effects?

\begin{figure}[ht!]
    \begin{center}
    \includegraphics[width=0.49\textwidth]{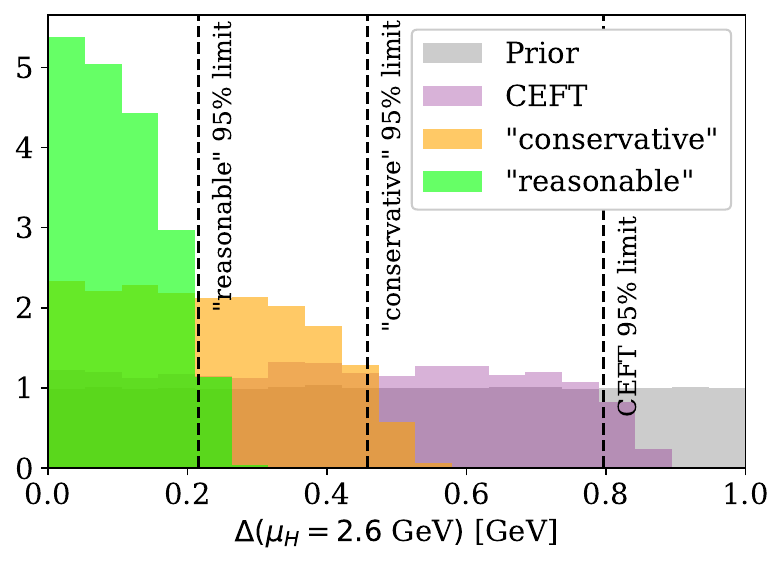}
    \end{center}
    \caption{Prior and posterior distributions for $\Delta$ at $\mu_H = 2.6$ GeV; see Sec.~\ref{sec:Bayesian} for the Bayesian analysis that yields these results. The purple posterior only uses information about the EoS at densities $n\leq 1.1\,n_s$ from CEFT calculations, with no input from astrophysical observations. The orange posterior uses the NS EoS inferred from astrophysical observations only up to the $\mu$ at the center of a 
    2.1~$M_\odot$ star, and requires only that $c_s^2 \leq 1$ at larger values of $\mu$. These are both overly conservative assumptions. The green posterior uses the inferred NS EoS  up to the $\mu$ at the center of the most massive stable NS that could be supported by a given EoS, and requires that $c_s^2 \leq \frac{1}{2}$ at larger values of $\mu$ than that. These are reasonable assumptions. The black dashed lines are at the 95\% upper credible limit for $\Delta$ for the three posterior distributions, with those for the reasonable/conservative assumptions, corresponding to $\Delta\leq 216/457$~MeV.}
    \label{fig:prior}
\end{figure}

Here we shall examine color superconductivity as one such candidate QCD effect. Above some density that is not reliably known but is around
the density of a neutron, neutrons and other baryons cannot exist, and matter 
is quark matter, with quarks of three colors and three flavors filling states with momenta up to their respective Fermi surfaces. At any density at which quark matter is found, Cooper pairs form at low enough temperatures because of the strong and attractive QCD interaction between pairs of quarks that are antisymmetric in color, making dense quark matter a color superconductor~\cite{Barrois:1977xd,Bailin:1983bm,Alford:1997zt,Rapp:1997zu,Alford:2007xm} with gaps $\Delta$ at all or some
Fermi surfaces. Pairing and the resulting gaps arise from the attraction between quarks, which is a leading-order feature of QCD, but the magnitude of $\Delta$ is nonperturbative in the QCD coupling.
In any color superconducting phase, 
the EoS receives nonperturbative contributions that are 
of order $\Delta^2\mu^2$, with $\mu$ the baryon number chemical potential.

At high enough densities, all nine quark Fermi momenta approach the same value $\mu/3$, and all nine quarks form Cooper pairs and are gapped, with a color-flavor locking (CFL) pairing pattern that is antisymmetric in flavor as well as color and that breaks separate color and flavor symmetries while leaving  symmetries that lock color and flavor rotations unbroken~\cite{Alford:1998mk}.

This most symmetric CFL pairing pattern is only possible if the $\Delta$ that results satisfies $\Delta \gtrsim m_s^2/(2\mu/3)$~\cite{Alford:2007xm}, 
where $m_s$ is the mass of the strange quark and where we are neglecting the up and down quark masses; this condition is very likely satisfied at the chemical potentials $\mu>2$~GeV of interest to us. 
The contribution to the pressure coming from CFL color-superconducting pairing is given by~\cite{Alford:1998mk,Alford:2004pf,Alford:2007xm}
\begin{equation}
p_{\rm CFL}=\frac{1}{3\pi^2}\Delta^2\mu^2
\label{pCFL}
\end{equation}
at zero temperature and, to a good approximation, at temperatures that are $\ll \Delta$.
The magnitude of $\Delta$ in the CFL phase has been calculated reliably only at the exceedingly high densities where the QCD coupling $g(\mu)$ is small. There, 
$\Delta/\mu \propto g^{-5} e^{-\frac{3\pi^2}{\sqrt{2} g}}$~\cite{Son:1998uk}, 
where the proportionality constant 
is given by 845.7~\cite{Schafer:1999jg,Pisarski:1999bf,Brown:1999aq,Wang:2001aq,Alford:2007xm}. 
The magnitude of $\Delta$ at densities that are not asymptotically large has been estimated in many models with estimates ranging between 20 and 250~MeV, most typically around 50-150~MeV~\cite{Alford:1997zt,Rapp:1997zu,Berges:1998rc,Alford:1998mk,Carter:1998ji,Rajagopal:2000wf,Alford:2007xm,Baym:2017whm,Leonhardt:2019fua,Braun:2021uua}, with one recent model calculation finding a value as large as 300~MeV~\cite{Braun:2021uua}. This supports our assumptions, that we shall rely upon throughout, that $T\ll \Delta$ in neutron stars  and that $\Delta$ is comfortably large enough for CFL pairing at the 
chemical potentials $\mu>2$~GeV we shall consider.
At lower densities, 
the separation between the Fermi momenta for quarks with differing flavors $\sim m_s^2/\mu$ 
forces the system into some less symmetric color superconducting phase for which $p_{\rm CSC}/(\Delta^2\mu^2) < 1/(3\pi^2)$~\cite{Alford:1999pa,Alford:2000ze,Bedaque:2001je,Kaplan:2001qk,Schafer:2001bq,Buballa:2001gj,Kundu:2001tt,Alford:2002rz,Alford:2003fq,Schmitt:2004et,Alford:2007xm,Anglani:2013gfu}.
We do not need to know anything about the pattern of pairing at these lower densities in order to derive the constraints that we shall describe. The impact on neutron star masses and radii of the modification of the pressure by color superconducting pairing at densities achieved in neutron stars has been investigated in model studies previously~\cite{Alford:2002rj,Alford:2004pf,Kurkela:2009gj,Steiner:2012xt}; our goal is different. We aim to use what is known about neutron stars to obtain model-agnostic constraints on the color-superconducting gap $\Delta$ at higher densities than are reached in neutron stars where the pQCD calculations we shall employ are well controlled and where quark matter is in the CFL phase.

As noted in Ref.~\cite{Zhou:2023zrm}, in previous model-agnostic Bayesian analyses seeking to use CEFT and/or pQCD calculations and/or astrophysical observations to constrain our knowledge of the dense matter EoS, the effects of color superconductivity on the EoS of the quark matter found at high  densities have been neglected, because these effects are suppressed by $\mathcal{O}(\Delta^2/\mu^2)$ relative to the dominant contributions to the pressure, which are $\mathcal{O}(\mu^4)$. (Specifically, $p_{\rm CFL}=p_0 (6\Delta/\mu)^2$, with $p_0$ the pressure of noninteracting quarks of three colors and flavors.)
Here, we remedy this neglect. The contributions to the pressure from CFL pairing are small as has long been understood, but we shall demonstrate that 
pQCD calculations and astrophysical measurements have now advanced to the point that  
we can begin using them in concert to obtain interesting constraints on the value of $\Delta$ in high density quark matter, 
and that will become only more so in the near future.
We shall see that adding $p_{\rm CFL}$ to the pressure of high density quark matter would already be inconsistent with present-day astrophysical observations if the value of $\Delta$ were to be at the high end of the range of published estimates.

\section{Analytic Estimate of the Constraint} \label{sec:analytic}
We start by deriving an analytic estimate for the maximal allowed value of the superconducting gap $\Delta$, employing knowledge about the EoS at low and intermediate densities from CEFT and neutron-star observations and at high densities from pQCD. This simple estimate is elevated to a full Bayesian analysis in the next Section, where the uncertainties of the neutron-star EoS inference as well as pQCD uncertainties are treated systematically.

Let us suppose that we know the EoS at low densities, $\mu\leq \mu_L$, and at high densities, $\mu \geq \mu_H$,
and seek an EoS interpolating between the lower-density thermodynamic point $(\mu_L,n_L,p_L)$ and the higher-density point $(\mu_H,n_H,p_H)$ that must (1) be mechanically stable, (2) be thermodynamically consistent, and (3) satisfy causality~\cite{Komoltsev:2021jzg}.
Consider a potential interpolating function for baryon number density $n(\mu)$ such that $n(\mu_L) = n_L$ and $n(\mu_H) = n_H$. 
The requirement of stability ensures that the interpolation is a monotonic single-valued function. Furthermore, the choice of $n(\mu)$ determines the pressure 
\begin{equation}
    p(\mu) = p_L+\int_{\mu_L}^\mu n(\mu) d\mu \,,
\end{equation}
and this choice is thermodynamically consistent only if $p(\mu_H) = p_H$. Last, the choice is causal if and only if 
\begin{equation}
    c_s^{-2} = \frac{\partial \log n}{\partial \log \mu} \geq 1,
\end{equation}
which imposes a minimum slope on $n(\mu)$. It is also possible to impose a stronger constraint on the speed of sound, $c_s^2 \leq c_{s,\rm max}^2$, resulting in a larger minimum slope of $n(\mu)$.

Given a speed of sound constraint $c_s^2 \leq c_{s,\rm max}^2$, we can draw the curve $n(\mu)$ between $(\mu_L,n_L,p_L)$ and $(\mu_H,n_H,p_H)$ which will result in the maximal possible final pressure $p_{\rm max}$: this maximal-pressure curve will extend from $n(\mu_H)=n_H$ down to $\mu_L$ with the minimum possible slope, and then undergo a phase transition at which it drops vertically to $n_L$. Similarly, 
we can identify
the minimal possible final pressure $p_{\rm min}$.
The resulting 
extremal
pressure differences between baryon chemical potentials $\mu_L$ and $\mu_H$ are
\begin{align}
    \delta p_{\rm max} &= \frac{n_H c_{s,\rm max}^2}{1+c_{s,\rm max}^2} \left(\mu_H - \mu_L \left(\frac{\mu_L}{\mu_H}\right)^{1/c_{s,\rm max}^2} \right) \,, \label{eq:pmax}\\
    \delta p_{\rm min} &= \frac{n_L c_{s,\rm max}^2}{1+c_{s,\rm max}^2} \left(\mu_H \left(\frac{\mu_H}{\mu_L}\right)^{1/c_{s,\rm max}^2} - \mu_L \right) \,. \label{eq:pmin}
\end{align}
If $p_H-p_L>\delta p_{\rm max}$ or $p_H-p_L<\delta p_{\rm min}$, no viable EoS can connect the endpoints of the low- and  high-density EoSs, $(\mu_L,n_L,p_L)$ and $(\mu_H,n_H,p_H)$. 

At high chemical potentials $\mu\geq \mu_H$, the matter is assumed to be in the CFL phase and the pressure and baryon density can be expressed as sums
\begin{align}
    p_H(\mu) &= p_{\rm pQCD}(\mu) + p_{\rm CFL}(\mu)\,, \label{pH-muH}\\
    n_H(\mu) &= n_{\rm pQCD}(\mu) + n_{\rm CFL}(\mu)\label{nH-muH}\,,
\end{align}
where $n_{\rm pQCD}$ and $p_{\rm pQCD}$ correspond to the perturbative contributions
and $n_{\rm CFL}$ and $p_{\rm CFL}$ are the contributions due to the presence of a color-superconducting gap. (Note
that Eqs.~\eqref{eq:pmax} and~\eqref{eq:pmin} could serve to constrain any effect modifying the pressure at high densities.)
For the perturbative contribution we take the current state-of-the-art N2LO perturbative QCD calculations at zero temperature with two massless quarks and strange quarks with ($\overline{\rm MS}$) mass $m_s$(2~GeV)=93.4~MeV~\cite{Kurkela:2009gj}
\footnote{We choose to use the highest-order calculation available that includes the strange-quark-mass dependence (see also \cite{Gorda:2021gha}). We make this choice even though a partial perturbative QCD calculation of the N3LO contribution to the pressure from massless quarks is available~\cite{Gorda:2021znl, Gorda:2023mkk} because $m_s^2$ and $\Delta^2$ appear similarly in thermodynamic potentials~\cite{Alford:2004pf}, which would make it inconsistent to include $\Delta$ without including $m_s$.}. At $\mu=\mu_H$, the superconducting contribution is given by Eq.~(\ref{pCFL}) evaluated at $\mu_H$ and
\begin{align}
        n_{\rm CFL}(\mu_H) &= \frac{2}{3\pi^2}\mu_H\,\Delta(\mu_H)^2 \,,
\end{align}
where here and throughout we have assumed that the gap is a slowly varying function of $\mu$.

When our goal is to constrain the gap, the maximum pressure difference of 
Eq.~(\ref{eq:pmax}) leads to a meaningful bound, whereas $\delta p_{min}$ is less relevant. 
Trusting the pQCD EoS above $\mu_H$ and a low energy equation of state (either NS EoS or CEFT) below $\mu_L$, we can set the actual pressure difference $p_H-p_L$ equal to the maximum pressure difference $\delta p_{\rm max}$ to extract the maximum possible superconducting gap at $\mu=\mu_H$, $\Delta(\mu_H)_{\rm max}$. Making the most conservative choice for the speed of sound, $c_{s,\textrm{max}}^2 =1$, leads to the compact expression
\begin{multline}
    \Delta_{\rm max}(\mu_H)^2 =  \frac{3\pi^2}{\mu_L^2} \Bigg[ \frac{n_{\rm pQCD}(\mu_H)}{2 \mu_H} \Bigl( \mu_H^2 - \mu^2_L \Bigr) \\
    - \Bigl( p_{\rm pQCD}(\mu_H)- p_L \Bigr) \Bigg].\label{AnalyticEstimate}
\end{multline}
For the purposes of demonstration, for now we choose $\mu_H = 2.6~\text{GeV}$, at which $n_{\rm pQCD}(\mu_H)\approx 40\,n_s$, a standard high density scale~\cite{Fraga:2013qra} where $p_{\rm pQCD}(\mu_H)\approx 3.6~{\rm GeV}/{\rm fm}^3$, and choose $(\mu_L,n_L,p_L)$ to reflect CEFT calculations at $n_L=1.1\,n_s$, which corresponds to choosing $\mu_L=0.97~\text{GeV}$~\cite{Hebeler:2013nza} where
$p_L \approx (2.2-3.5)~{\rm MeV}/{\rm fm}^3$.
With these choices for $(\mu_L,n_L,p_L)$ and $(\mu_H,n_H,p_H)$, we can use Eq.~(\ref{AnalyticEstimate}) to estimate $\Delta_{\rm max} (\mu_H = 2.6~\text{GeV})\approx 880~\text{MeV}$. 
If instead we choose $\mu_L=1.45$~GeV and $p_L = 160~{\rm MeV}/{\rm fm}^3$, which are 
reasonable 
estimates for the chemical potential and pressure at the center of a 
$2.1~M_\odot$ neutron star, we estimate $\Delta_{\rm max} (\mu_H = 2.6~\text{GeV})\approx 440~\text{MeV}$.
In the following Section, we will make this constraint on $\Delta$ precise by systematically treating the uncertainties in the astrophysical observations and in the pQCD calculations in a complete Bayesian analysis and additionally find stronger constraints by imposing reasonable stricter limits on the speed of sound at the highest densities.

\section{Bayesian Constraint on the Gap}
\label{sec:Bayesian}
We will now study the posterior distribution of the gap $\Delta$ constrained by the astrophysical observations, denoted as $P(\Delta | \data)$. The gap is not directly constrained by neutron-star observations, but the observations constrain the EoS and therefore the values of $(\mu_L , n_L, p_L)$ we may employ in Eqs.(\ref{eq:pmax}) and (\ref{eq:pmin}). 
EoS inference gives access to the 
posterior distribution for the EoS at neutron star densities as constrained by astrophysical data, $P({\rm EoS}| \data)$; 
here, we will use the posterior distribution from Ref.~\cite{Gorda:2022jvk} that infers the EoS incorporating the mass-measurements of PSR J0348+0432~\cite{Antoniadis:2013pzd} and PSR J1624-2230~\cite{Fonseca:2016tux}, the simultaneous mass and radius measurement of PSR J0740+6620 obtained using the NICER telescope~\cite{Miller:2021qha} as well as the tidal deformability measurement of GW170817 achieved by the LIGO/Virgo collaboration~\cite{LIGOScientific:2018hze}. 
In addition, the electromagnetic counterpart of GW170807 is accounted for by assuming that the final merger product is a black hole.
This posterior consists of a large sample of EoSs generated using a Gaussian process prior conditioned with the CEFT up to $1.1\,n_s$~\cite{Hebeler:2013nza} and the above mentioned NS observational data.

The posterior distribution for $\Delta$ can then be obtained by summing over all the members of the EoS sample so as to marginalize over the inferred EoS,  
\begin{align}
    P(\Delta | \data) & = \int_{\EoS} P(\Delta | \EoS ) P(\EoS | \data ).
\end{align}
The conditional probability for the gap given the EoS $P(\Delta | \EoS )$, can be inferred using Bayes' theorem
\begin{align}
   P(\Delta | \EoS ) = \frac{P(\EoS | \Delta ) P(\Delta)}{P(\EoS)}
\end{align}
for a given prior distribution for the gap $P(\Delta)$. The likelihood function $P(\EoS | \Delta )$ describes the probability of the low-density EoS given the model parameter $\Delta$. Following Ref.~\cite{Komoltsev:2021jzg}, the likelihood function is taken to be unity if the condition $\delta p_{min} < p_H - p_L < \delta p_{max}$ is satisfied, otherwise the function is taken to be zero. 

For the high-density EoS at (and above) $\mu=\mu_H$,
we use $n_H(\mu)$ and $p_H(\mu)$ as defined in Eqs.~(\ref{pH-muH}) and (\ref{nH-muH}), that is, as a sum of the perturbative result and the gap contribution. The perturbative renormalization scale uncertainty is accounted for by marginalizing over the renormalization scale parameter $X\in [1/2, 2]$ in the scale-averaging 
prescription~\cite{Gorda:2023usm} as in Ref.~\cite{Gorda:2022jvk}. We must also choose the $\mu$ at which we terminate our inferred EoS (taken from the posterior in Ref.~\cite{Gorda:2022jvk}); below this $\mu=\mu_L$, we use the inferred EoS; above $\mu_L$, we employ the analytical argument of Sect.~\ref{sec:analytic} with $\mu_L$ given by this $\mu$.

\begin{figure}
\begin{center}
\includegraphics[width=.48\textwidth]{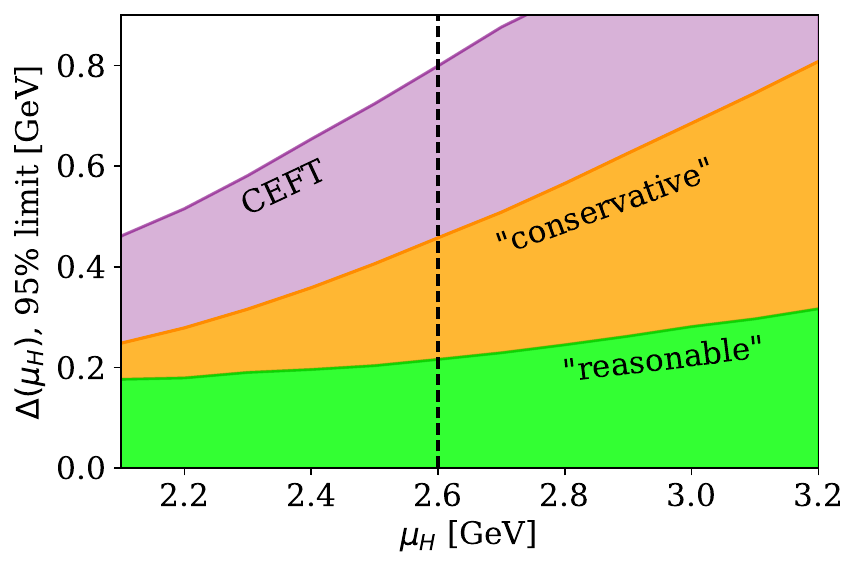}
\end{center}
\caption{The upper boundary of each colored region shows the 95\%-credible upper limit on the color-superconducting gap $\Delta(\mu_H)$ as a function of baryon number chemical potential $\mu_H$ extracted in three different ways, all using a uniform prior for $\Delta$ in the range $0 \leq \Delta \leq 1~\text{GeV}$. The dashed vertical line indicates $\mu_H = 2.6~\text{GeV}$, the standard scale at which pQCD uncertainties are  well under control~\cite{Fraga:2013qra} at which 
we have plotted Fig.~\ref{fig:prior}. The 
boundary of the purple 
region indicates the constraint on $\Delta$ obtained using only CEFT up to $1.1\,n_s$ and the logic of Sect.~\ref{sec:analytic} between there and $\mu_H$. The 
boundaries of the orange and green regions indicate the upper limits on $\Delta$ found using the astrophysically inferred EoS posterior either by extending the prior to the central density of the heaviest possible stable NS  and applying the logic of Sect.~\ref{sec:analytic} with the assumption that $c_s^2 <1/2$ at still higher densities (``reasonable") or, more conservatively, by using the posterior no further than the central density of a $2.1~M_\odot$ NS and requiring only causality (``conservative") at higher densities. }\label{fig:cetvsastro}
\end{figure}

The first thing we can try is to choose $n_L=1.1\,n_s$ and $\mu_L=0.97$~GeV, below which the EoS is determined by CEFT, and above this $\mu_L$ employ the argument of Sect.~\ref{sec:analytic} while requiring only causality, $c_s^2\leq 1$. This means that we are not including anything we know from astrophysical observations about the EoS at the densities above $1.1\,n_s$ 
found in neutron stars. 
 Fig.~\ref{fig:prior} displays the resulting posterior distribution for $\Delta$ at $\mu_H = 2.6$~GeV obtained from this analysis (labelled as "CEFT") showing that the gap is consistent with zero, but the highest values $\Delta$ are excluded. Quantifying the upper limit through the 95th percentile of the distribution  $P(\Delta | {\rm data})$ gives us an upper limit of $799~\textrm{MeV}$, which is in expected agreement with the analytic estimate presented in Sect.~\ref{sec:analytic}. 

The maximally conservative approach 
which uses
both CEFT and astrophysical observations is to set $\mu_L$ to be the chemical potential reached at the center of a 2.1~$M_\odot$ NS~\footnote{ Assuming that this is less than $\mu_{\rm TOV}$; for EoSs taken from the posterior ensemble for which this is not the case, we use $\mu_{\rm TOV}$. Note that our choice of termination density is different from that in Ref.~\cite{Gorda:2022jvk}, where the aim was to infer the EoS up to a given density. For a detailed discussion of the termination density, see Ref.~\cite{Komoltsev:2023zor}.}, and above 
this $\mu_L$ to only require  $c_s^2\leq 1$ in the argument of Sect.~\ref{sec:analytic}. 
In the following we will call this the ``conservative'' scenario.

A reasonable approach 
to incorporating what we know is to push $\mu_L$ somewhat higher, to its value at the center of the most massive stable neutron star supported by a given EoS taken from the posterior ensemble, leading to $(\mu_L, n_L, p_L) = (\mu_{\rm TOV}, n_{\rm TOV}, p_{\rm TOV})$, where the subscript ${\rm TOV}$ refers to 
this maximal density that could be realized 
in stable neutron stars. 
Furthermore, while the most conservative speed of sound constraint we can place on the EoS at chemical potentials $\mu>\mu_L$ is causality ($c_s^2 \leq 1$), if we take $\mu_L=\mu_{\rm TOV}$ this is high enough that a more reasonable constraint on $c_s^2$ at $\mu>\mu_L$ should be somewhat lower. 
While the existence of $M\gtrsim 2 M_\odot$ stars suggests that the sound speed may exceed the conformal value of $c_s^2 = 1/3$ at densities reached in stable neutron stars~\cite{Bedaque:2014sqa}, 
we know from pQCD calculations that at higher densities, say for $2.2~{\rm GeV} < \mu < 2.6~{\rm GeV}=\mu_H$, $c_s^2$ is close to 1/3 and certainly  below 1/2~\cite{Komoltsev:2023zor, Gorda:2022jvk, Annala:2023cwx}.
Furthermore, 
following an argument from  Ref.~\cite{Komoltsev:2023zor} and Sect.~\ref{sec:analytic}, 
if we require $c_s^2\leq 1$ for $\mu_L<\mu<\mu_H$ 
then the maximal-pressure EoS that serves to set the constraint on $\Delta$ will have $c_s^2=1$ over a wide range of $\mu$ on the high side of this range, up to $\mu_H$, in sharp disagreement with what we know from pQCD.
Furthermore, at the highest densities the sound speed approaches the perturbative limit (with gap $\Delta$) 
{\it from below}. It is therefore unlikely that the sound speed reaches values anywhere near unity in the density range $n_{\rm TOV} < n < n_H$. 
In the following, we therefore define a ``reasonable'' scenario in which we extend the prior to the end of stability, setting $\mu_L=\mu_{\rm TOV}$, and require $c_s^2 < 1/2$ beyond.

As can be seen from Fig.~\ref{fig:prior}, the values of the gap, $\Delta(\mu_H = 2.6~\textrm{GeV})$, become more constrained  upon including astrophysical information, while the distributions for both scenarios are still consistent with vanishing gap. For the conservative scenario we find a 95\% upper limit of 457~MeV, while for the reasonable scenario the upper limit is 216~MeV. 
We see that the 
simpler analytic estimate of Sect.~\ref{sec:analytic} yielded an upper limit on $\Delta$ in rough agreement with what we can now conclude more reliably from our conservative Bayesian analysis.
The chemical potential dependence of the conservative and reasonable upper limits are displayed in 
Fig.~\ref{fig:cetvsastro}, with the CEFT results at $n_L=1.1\,n_s$ discussed above shown also, for comparison. The upper limit on $\Delta(\mu_H)$ becomes tighter at lower densities, with the caveat that the bound starts to become less reliable given the poorer convergence of the pQCD calculation at lower values of $\mu_H$. 

Although it is (at least to us) already impressive that values of $\Delta(\mu_H)$ above $\sim 200$~MeV (meaning those on the high side of prior theoretical expectations) are disfavored by what we know from astrophysical observations and pQCD calculations, the most important implication of our work is that future observations
and future advances in pQCD calculations will further tighten the constraints on $\Delta(\mu_H)$. We explore some possibilities in Appendix~\ref{sec:appendix}.

\section{Looking Ahead}

Although careful model studies have been employed to investigate the impact of color superconductivity on the equation of state at neutron star densities, see e.g.~Refs.~\cite{Alford:2002rj,Alford:2004pf,Kurkela:2009gj,Steiner:2012xt},
it has long been assumed that the EoS is not sufficiently sensitive to quark pairing for the physics of color superconductivity to be probed by studying the EoS using model-agnostic methods~\cite{Alford:2007xm}. 
We have shown that the combination of what we know about the EoS from neutron-star observations and from pQCD calculations already places a meaningful model-independent 95\% credible upper bound on the CFL gap $\Delta(2.6~\mathrm{GeV})< 216~\mathrm{MeV}$ that challenges some of the higher model predictions for this fundamental quantity.

Because the gap adds to the pressure, it makes it more difficult to satisfy the integral condition $p_L > p_H - \delta p_{\rm max}$. The implication of this for using the QCD calculation in EoS inference is that adding the gap makes the constraint on the EoS stronger. Therefore, leaving it out (as has been done prior to this work) is a conservative choice.
That the increase in pressure due to a moderate gap $\Delta$ can make the QCD EoS inconsistent with present astrophysical observations is a demonstration of the power of the integral constraints, in concert with today's astrophysical data.

The strength of the upper limit is directly linked to the pQCD calculation, providing a strong motivation to enhance its accuracy. 
The N2LO results for the pQCD EoS have a substantial scale-variation uncertainty\cite{Kurkela:2014vha,Gorda:2023usm}, which is anticipated to decrease significantly upon the completion of the N3LO calculation that is currently in progress~\cite{Gorda:2018gpy, Gorda:2021znl,Gorda:2021kme,Gorda:2023mkk}. 
Depending on where the N3LO result for the pressure lands, together with astrophysical observations it could yield a more stringent upper bound on $\Delta$ or, conceivably, could favor nonzero values
of this quantity.
In the previous Section, and in more detail in the Supplemental Material, we have provided examples of how hypothetical future astrophysical measurements of neutron star masses, radii, and tidal deformabilities could tighten the constraint on $\Delta$; these considerations only add to the already strong motivation for making these measurements.

Finally, the determination of the gap has astrophysical consequences that go beyond 
the EoS~\cite{Alford:2007xm}. The magnitude of the gap $\Delta$, together with the pairing pattern (which depends on $\mu$ and $m_s$ as well as on $\Delta$), can have significant effects on transport properties in quark matter and may result in observable consequences. These may include variations in the cooling rates of neutron stars due to changes in their heat capacity, emissivity, and thermal conductivity as well as effects on the ringdowns of mergers and damping of r-modes caused by changes in bulk and/or shear viscosity.

\

We acknowledge helpful conversations with Mark Alford, Tyler Gorda, Jamie Karthein, Tore Kleppe, Oleg Komoltsev, Joonas N\"attil\"a, Bruno Scheihing-Hitschfeld and Andreas Schmitt. RS gratefully acknowledges the hospitality of the CERN Theory Department. The work of KR and RS is supported by the U.S.~Department of Energy, Office of Science, Office of Nuclear Physics grant DE-SC0011090.

\appendix
\section{Possible Consequences  of Future Measurements} \label{sec:appendix}

\begin{figure}
    \begin{center}
    \includegraphics[width=.49\textwidth]{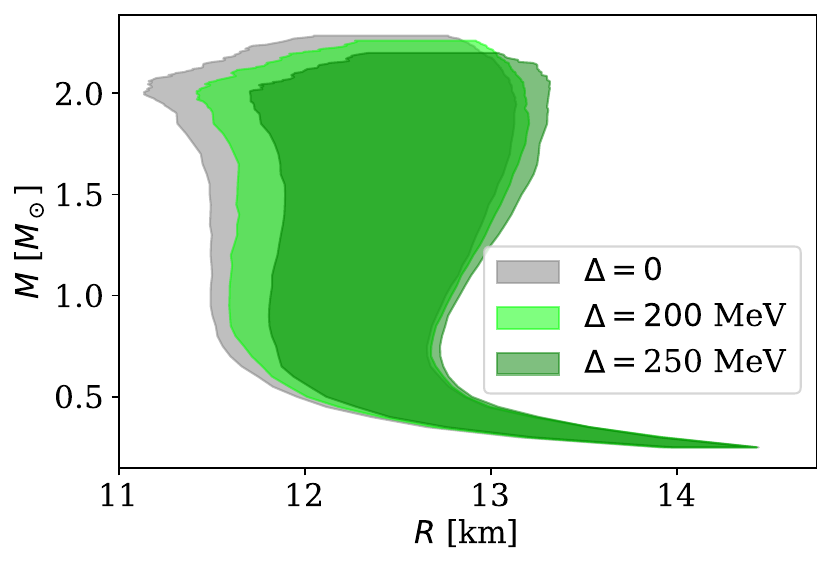}
    \end{center}
    \caption{5\%-95\% boundaries of the posterior distributions for the radii of NS with a given mass, assuming color superconducting gaps of 0, 200~MeV, and 250~MeV at $\mu_H=2.6$~GeV in the ``reasonable" scenario described in the text. Larger values of the gap can be disfavored (favored) by the discovery of NS with small (large) radii; they also imply a preference for slightly smaller maximum stable NS masses.} \label{fig:mrplot}
\end{figure}

We anticipate that future astrophysical measurements and future advances in pQCD calculations will further tighten the constrains on $\Delta(\mu_H)$.
To get a sense of this, in Fig.~\ref{fig:mrplot} we show how the 5\%-95\% credible interval for the radii of neutron stars with a given mass obtained from our posterior distribution of EoSs in the reasonable scenario varies depending on whether we posit $\Delta(2.6~{\rm GeV})=0$, 200 or 250~MeV. This plot indicates that future measurements that favor small (large) radii will tend to disfavor (favor) larger values of the color superconducting gap. For example, we have checked that a hypothetical measurement of a $1.4~M_{\odot}$ NS with a radius of $11.6\pm 0.1$~km ($13.1\pm 0.1$~km) would tighten the upper bound on $\Delta(2.6~{\rm GeV)}$ from the current 216~MeV to 203~MeV (loosen the upper bound to 232~MeV) while tilting the histogram in Fig.~\ref{fig:prior} toward smaller (larger) values of $\Delta$. 
A hypothetical future measurement of a 
$2.2~M_{\odot}$ NS with a radius of $12\pm 0.1$~km would yield a 95\% credible upper limit 
of $\Delta(2.6~{\rm GeV})<198$~MeV.

\begin{figure*}
    \begin{center}
    \includegraphics[width=1.00\textwidth]{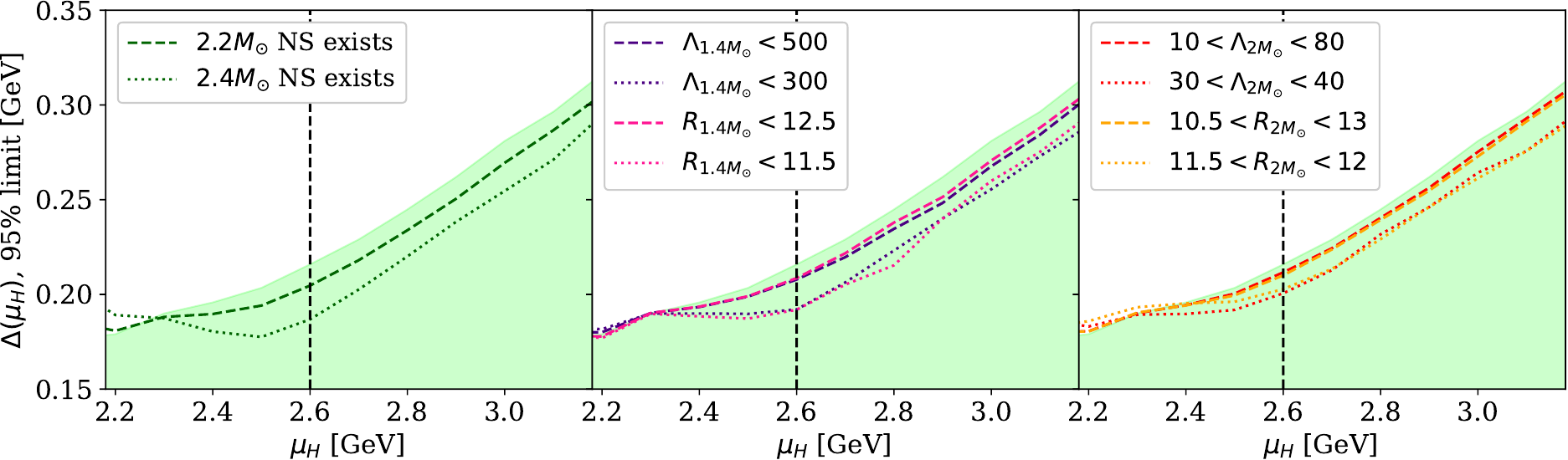}
    \end{center}
    \caption{Impact of hypothetical future measurements of NS masses, radii, or tidal deformabilities on the constraint on the color superconducting gap obtained via our ``reasonable'' Bayesian analysis. In each panel, the light green shaded region corresponds to the 95\% credible upper limit found using current astrophysical data and a bound of $c_s^2 \leq \frac{1}{2}$, as shown in green in Figs.~\ref{fig:prior} and \ref{fig:cetvsastro}. Each curve shows the impact on the bound of adding one additional hypothetical astrophysical measurement. These curves are obtained by considering only those EoSs in the posterior ensemble which satisfy the hypothetical measurement. In general, larger maximum masses, smaller radii, and smaller tidal deformabilities result in a tighter upper limit on $\Delta(\mu_H$).} \label{fig:singlemeasure}
\end{figure*}

\begin{figure*}
    \begin{center}
    \includegraphics[width=0.7\textwidth]{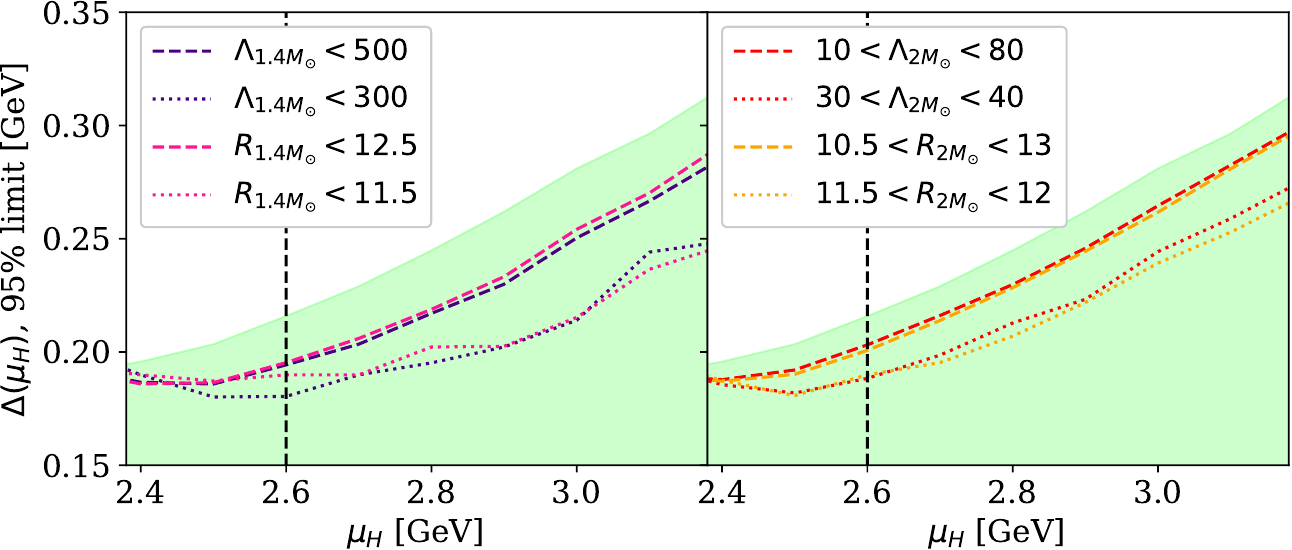}
    \end{center}
    \caption{Impact of a combination of two of the hypothetical future measurements from Fig.~\ref{fig:singlemeasure} on the upper limit on $\Delta(\mu_H)$. Each curve shows the effect of a measurement showing the existence of a $2.2\,M_\odot$ NS together with a second measurement constraining either the tidal deformability or radius of a $1.4\,M_\odot$ neutron star (left panel) or a $2.0\,M_\odot$ neutron star (right panel).
    The effect of the observation of a $2.2\,M_\odot$ NS together with any one of the second hypothetical measurements makes the upper limit on $\Delta(\mu_H)$ more stringent than in the cases plotted in Fig.~\ref{fig:singlemeasure}. }\label{fig:combomeasure}
\end{figure*}

We conclude this Appendix with a look at how hypothetical future astrophysical measurements 
could change the posterior constraint on the CFL superconducting gap $\Delta(\mu_H)$ that we obtain from our Bayesian analysis incorporating those measurements together with today's astrophysical observations, CEFT calculations, and pQCD calculations. Figs.~\ref{fig:singlemeasure} and~\ref{fig:combomeasure} show how introducing various hypothetical upper/lower bounds on NS radii, masses, or tidal deformability into our Bayesian analysis of the equation of state would each bring down the 95\% credible upper limit on the gap $\Delta(\mu_H)$.
We observe in Fig.~\ref{fig:singlemeasure} that the discovery of pulsars with masses reliably determined to be at or above $2.2\,M_\odot$, or the discovery of an upper limit on the radius of a $M=1.4 \,M_\odot$ neutron star $R_{1.4 \odot}< 12$~km (or the discovery of a corresponding limit on the tidal deformability of $\Lambda_{1.4 \odot}\lesssim 400$) would tighten the upper bound on $\Delta(\mu_H)$ by about 10-20~MeV. Note that this strengthened constraint may already be implied by the updated radius measurement of PSR J0030+0451 reported recently by NICER \cite{Vinciguerra:2023qxq}. Similar tightening of the upper bound are possible if a $M=2.0 \,M_\odot$ neutron star were observed to have a radius in the range $11.5~{\rm km} < R_{2.0 \odot} < 12~{\rm km}$ (or if its tidal deformability were observed to be in the corresponding range  $30 \Lambda_{2.0 \odot}\lesssim 40$). Fig.~\ref{fig:combomeasure} highlights that the consequence of the discovery of pulsars with masses $\geq 2.2\,M_\odot$ together with one additional hypothetical future measurement could yield an even more stringent limit on $\Delta(\mu_H)$, tighter by about 20-40~MeV.  At $\mu_H=2.6$~GeV, all of these conclusions are in accord with what we were able to conclude more qualitatively from Fig.~\ref{fig:mrplot}. 
We have plotted Fig.~\ref{fig:singlemeasure} only for $\mu_H>2.2$~GeV and Fig.~\ref{fig:combomeasure} only for $\mu_H>2.4$~GeV because we find that the statistical power of the ensemble that we are using to draw inferences weakens at lower $\mu_H$ as we add each additional hypothetical measurement.

\bibliography{main}

\end{document}